# CSIndexbr: Exploring the Brazilian Scientific Production in Computer Science

Marco Tulio Valente
Department of Computer Science
Federal University of Minas Gerais
mtov@dcc.ufmg.br

Klerisson Paixao
Faculty of Computing
Federal University of Uberlandia
klerisson@ufu.br

**Abstract**

CSIndexbr is a web-based system that provides meaningful, open, and transparent data about Brazilian scientific production in Computer Science. Currently, the system collects full research papers published in the main track of selected conferences. The papers are retrieved from DBLP. In this article, we describe the main features and resources provided by CSIndexbr. We also comment on how other researchers can use the data provided by the system to analyze the Brazilian production in Computer Science.

## 1. Introduction

CSIndexbr (http://csindexbr.org) is a web-based system that provides meaningful, open, and transparent data about Brazilian scientific production in Computer Science. Currently, the focus is on full research papers published in the main track of selected conferences. The papers are retrieved from DBLP (http://dblp.org), which is a repository providing meta-data about scientific publications in Computer Science.

CSIndexbr is maintained by the Applied Software Engineering Research Group, from Federal University of Minas Gerais, Brazil. However, the project results are not endorsed by UFMG or any Brazilian research agency.

The source code of the system is publicly available on GitHub, under a MIT license; the data collected by the system is also available on GitHub, under a Creative Commons CC BY-NC-SA 4.0. This license grants free and non-commercial usage of the licensed material, including adaptations. However, proper credit should be granted, by linking to CSIndexbr site and/or citing this paper.

## 2. Scope

### 2.1 Research Areas

CSIndexbr tracks papers published in several sub-areas of Computer Science. Currently, the system covers 18 sub-areas: Software Engineering, Programming Languages, Information Systems, Human-Computer Interaction, Computer Networks, Mobile Computing, Distributed Systems, Computer Architecture & HPC, Databases, Web & Information Retrieval, Data Mining & Machine Learning, Artificial Intelligence, Algorithms & Complexity, Security & Cryptography, Computer Vision, Formal Methods & Logic, Robotics, and Computer Science Education.

## 2.2 Venues

CSIndexbr indexes full papers published in the main track of selected Computer Science Conferences. In other words, the focus is on papers describing mature and carefully evaluated work. Papers tracked by CSIndexbr should be viewed as journal-quality papers, although published in conferences. For this reason, the system does not indexes short papers, including demos, tool papers, early research papers, papers in industry track, etc.

The conferences tracked by CSIndexbr should attend the following criteria:

CSIndexbr conferences $\Rightarrow$ *submitted* > 100 papers and *acceptance* < 30% and *h5-index* > 20.

We consider the 2017 edition of the conferences to collect the number of submitted and accepted papers. H5-index is is the largest number *h* such that *h* articles published *in the last five editions* of a conference have at least *h* citations each. We used h5-index as computer by Google Scholar, in the first semester of 2017. Finally, most selected conferences have well-known sponsors, such as as ACM SIGs, IEEE CS, etc.

However, in all sub-area there a few conferences that only attend partially the proposed criteria. Since they are important conferences, we decided to index them. The metrics that are not attended by these conferences are presented under a yellow color in the *stats* page of the system.

For example, Figure 1 shows the Software Engineering conferences tracked by CSIndexbr. As we can see, three conferences do not attend the minimal number of 100 submissions in 2017 (MODELS, SPLC, and ICPC); furthermore, two conferences have acceptance rates slightly greater than 30% (ISSRE and ICPC) and one conference has a h5-index less than 20 (ICSA).

| | Conference | Sponsor | Submitted | Accepted | Accept. Rate | h5-index | Rank | Pages |
|---|---|---|---|---|---|---|---|---|
| 1 | ICSE | ACM SIGSOFT/IEEE CS | 415 | 68 | 16.4 | 68 | top | 12 |
| 2 | FSE | ACM SIGSOFT | 295 | 72 | 24.4 | 43 | top | 12 |
| 3 | ASE | ACM SIGSOFT/IEEE CS | 314 | 65 | 20.7 | 31 | near-the-top | 12 |
| 4 | MSR | ACM SIGSOFT/IEEE CS | 121 | 37 | 30.6 | 39 | | 12 |
| 5 | ISSTA | ACM SIGSOFT | 118 | 31 | 26.3 | 31 | | 12 |
| 6 | ICSME | IEEE CS | 150 | 42 | 28 | 29 | | 12 |
| 7 | ICST | IEEE CS | 135 | 36 | 26.7 | 29 | | 12 |
| 8 | MODELS | ACM SIGSOFT/IEEE CS | 68 | 17 | 25 | 26 | | 11 |
| 9 | SANER | IEEE CS | 135 | 34 | 25.2 | 26 | | 12 |
| 10 | SPLC | - | 49 | 15 | 30.6 | 25 | | 10 |
| 11 | RE | IEEE CS | 96 | 27 | 28.1 | 23 | | 10 |
| 12 | FASE | ETAPS | 91 | 25 | 27.5 | 23 | | 17 |
| 13 | ISSRE | IEEE CS | 109 | 34 | 31.5 | 22 | | 12 |
| 14 | ICPC | IEEE CS | 83 | 28 | 33.7 | 21 | | 12 |
| 15 | ESEM | ACM SIGSOFT/IEEE CS | 109 | 21 | 19.3 | 20 | | 10 |
| 16 | ICSA | IEEE | 95 | 21 | 22.1 | 16 | | 10 |

Fig. 1: Statistics about Software Engineering Conferences (values in yellow are exceptions to thresholds used by CSIndexbr)

CSIndexbr also classifies some conferences as *top-conferences*. To select the top-conferences, ee combined an initial metric-based assessment with a qualitative judgment. First, top conferences should attend the following criteria:

CSIndexbr top-conferences ⇒ *submitted* > 180 and *h5-index* > 40.

Next, we manually evaluated the conferences selected in each area. We checked whether each candidate conference T from an area A is distinct from the other conferences in A; we also require T to cover a broad range of topics in A, instead of being a specialized conference. We selected at most two top-conferences per area, with exceptions in the case of areas that congregate distinct sub-areas, such as Computer Architecture & High Performance Computing. Finally, we classify as near-the-top the conferences with metrics similar to the top-conferences of their areas.

## 2.3 Researchers and Departments

CSIndexbr tracks papers published by around 900 Brazilian professors. These professors are affiliated to Computer Science departments of federal universities (UFs), state universities, private universities (mostly, PUCs) or of technological institutes (IFs and CEFETs).

The systems also computes a score for the CS departments, as follows:

*department-score* = *A* + (0.66 * *B*) + (0.33 * *C*)

where *A* is the number of papers in top conferences; *B* is the number of papers in "near-the-top" conferences; and *C* is the number of papers in the remaining conferences listed by CSIndexbr.

**3. Data**

For each research area, CSIndexbr provides the following information:

1. Number of full papers published by Brazilian professors in the selected conferences, in the last five years;
2. Department scores, considering their production in a given area.
3. The number of professors in each Brazilian department with papers in the selected conferences.
4. Meta-data about the papers published in a research area, including title, authors, authors affiliation and DOI.

It is also possible to retrieve all papers published by a given Brazilian professor, in the conferences tracked by CSIndexbr.